\documentclass[twocolumn]{aastex701}
\usepackage{graphicx}%
\usepackage{amsmath,amssymb,amsfonts}%
\usepackage{amsthm}%
\usepackage{mathrsfs}%
\usepackage[title]{appendix}%
\usepackage{xcolor}%
\usepackage{comment}
\usepackage{caption}
\usepackage{tabularray}
\usepackage{hyperref}

\newcommand{\lam}{$\lambda$}

\newcommand{\halpha}{$\mathrm{H}{\alpha}$}
\newcommand{\hbeta}{$\mathrm{H}{\beta}$}
\newcommand{\hgamma}{$\mathrm{H}{\gamma}$}

\newcommand{\nii}{\hbox{[N\,{\sc ii}]}} 
 
\newcommand{\jwst}{{\it JWST}}
\newcommand{\oiii}{\hbox{\sc [O\,iii]}}     

\newcommand{\neiii}{\hbox{[Ne\,{\sc iii}]}}

\newcommand{\oiiiauroral}{[\textrm{O}\,\textsc{iii}]\ensuremath{\lambda4364}}
\newcommand{\oiiidbl}{[\textrm{O}\,\textsc{iii}]\ensuremath{\lambda\lambda4960,5008}}

\newcommand{\fevii}{[\textrm{Fe}\,\textsc{vii}]}

\newcommand{\feii}{[\textrm{Fe}\,\textsc{ii}]}
\newcommand{\feiipermitted}{\textrm{Fe}\,\textsc{ii}}

\begin{document}

\title{The GlimmIr: Spectroscopic Variability in a z$\sim$7 LRD Indicates\\ Rapid Changes in Both the Narrow and Broad Line Regions}

\suppressAffiliations

\author[orcid=0000-0000-0000-0001,sname='Lambrides']{Erini Lambrides}
\affiliation{Astrophysics Science Division, NASA Goddard Space Flight Center, 8800 Greenbelt Rd, Greenbelt, MD 20771, USA}
\affiliation{Department of Astronomy, University of Maryland, College Park, MD 20742, USA}
\affiliation{Center for Research and Exploration in Space Science and Technology, NASA/GSFC, Greenbelt, MD 20771 USA}
\email[show]{erini.lambrides@nasa.gov}

\author[orcid=0000-0001-6251-4988,sname='Hutchison']{Taylor A. Hutchison} 
\altaffiliation{NASA Postdoctoral Fellow}
\affiliation{Astrophysics Science Division, NASA Goddard Space Flight Center, 8800 Greenbelt Rd, Greenbelt, MD 20771, USA}
\affiliation{Department of Astronomy, University of Maryland, Baltimore Country, MD 21250, USA}
\affiliation{Center for Research and Exploration in Space Science and Technology, NASA/GSFC, Greenbelt, MD 20771 USA}
\email{taylor.hutchison@nasa.gov}

\author[orcid=0000-0003-2366-8858,sname='Larson']{Rebecca L.\ Larson}
\altaffiliation{Giacconi Postdoctoral Fellow}
\affil{Space Telescope Science Institute, 3700 San Martin Drive, Baltimore, MD 21218, USA}
\email{rlarson@stsci.edu}

\author[orcid=0000-0002-7959-8783,sname='Arrabal Haro']{Pablo Arrabal Haro}
\affiliation{Center for Space Sciences and Technology, UMBC, 5523 Research Park Dr, Baltimore, MD 21228 USA }
\affiliation{Astrophysics Science Division, NASA Goddard Space Flight Center, 8800 Greenbelt Rd, Greenbelt, MD 20771, USA}
\email{pablo.arrabalharo@nasa.gov}

\author[orcid=0000-0001-7503-8482,sname='Papovich']{Casey Papovich}
\affiliation{Department of Physics and Astronomy, Texas A\&M University, College Station, TX, 77843-4242 USA}
\affiliation{George P. and Cynthia Woods Mitchell Institute for Fundamental Physics and Astronomy,\\ Texas A\&M University, College Station, TX, 77843-4242 USA}
\email{papovich@tamu.edu} 

\author[orcid=0000-0000-0000-0001,sname='Hu']{Weida Hu}
\affiliation{Department of Physics and Astronomy, Texas A\&M University, College Station, TX, 77843-4242 USA}
\affiliation{George P. and Cynthia Woods Mitchell Institute for Fundamental Physics and Astronomy,\\ Texas A\&M University, College Station, TX, 77843-4242 USA}
\email{weidahu@tamu.com}  

\author[orcid=0000-0001-7151-009X,sname='Cleri']{Nikko J.\ Cleri}
\affiliation{Department of Astronomy and Astrophysics, The Pennsylvania State University, University Park, PA 16802, USA}
\affiliation{Institute for Computational and Data Sciences, The Pennsylvania State University, University Park, PA 16802, USA}
\affiliation{Institute for Gravitation and the Cosmos, The Pennsylvania State University, University Park, PA 16802, USA}
\email{cleri@psu.edu}

\author[0000-0001-8519-1130]{Steven L. Finkelstein}
\affiliation{Department of Astronomy, The University of Texas at Austin, Austin, TX, USA}
\affiliation{Cosmic Frontier Center, The University of Texas at Austin, Austin, TX, USA}
\email{stevenf@astro.as.utexas.edu}

\author[0000-0002-1410-0470]{Jonathan R. Trump}
\affil{Department of Physics, 196A Auditorium Road, Unit 3046, University of Connecticut, Storrs, CT 06269, USA}
\email{jonathan.trump@uconn.edu}

\author[0000-0003-4528-5639]{Pablo G. P\'erez-Gonz\'alez}
\affiliation{Centro de Astrobiolog\'{\i}a (CAB), CSIC-INTA, Ctra. de Ajalvir km 4, Torrej\'on de Ardoz, E-28850, Madrid, Spain}
\email{pgperez@cab.inta.csic.es}

\author[0000-0001-9269-5046]{Bingjie Wang}
\thanks{NHFP Hubble Fellow}
\affiliation{Department of Astrophysical Sciences, Princeton University, Princeton, NJ 08544, USA}
\email{bjwang@princeton.edu}

\author[0000-0002-8360-3880]{Dale D. Kocevski}
\affiliation{Department of Physics and Astronomy, Colby College, Waterville, ME 04901, USA}
\email{}

\author[0000-0002-0302-2577]{John Chisholm}
\affiliation{Department of Astronomy, The University of Texas at Austin, Austin, TX 78712}
\affiliation{Cosmic Frontier Center, The University of Texas at Austin, Austin, TX 78712}
\email{chisholm@austin.utexas.edu}

\author[0000-0002-1174-2873]{Amy Secunda}
\email{asecunda@flatironinstitute.org}
\affiliation{Center for Computational Astrophysics, Flatiron Institute, 162 Fifth Avenue, New York, NY 10010, USA}

\author[0000-0001-8582-7012]{Sarah E.~I.~Bosman}
\affiliation{Institute for Theoretical Physics, Heidelberg University, Philosophenweg 12, D–69120, Heidelberg, Germany}
\affiliation{Max-Planck-Institut f\"{u}r Astronomie, K\"{o}nigstuhl 17, 69117 Heidelberg, Germany}
\email{bosman@thphys.uni-heidelberg.de}

\author[0000-0003-3596-8794]{Hollis Akins}
\affiliation{Department of Astronomy, The University of Texas at Austin, Austin, TX, USA}
\email{hollis.akins@utexas.edu}

\author[0000-0003-2495-8670]{Mitchell Karmen}
\affiliation{The William H. Miller III Department of Physics \& Astronomy, Johns Hopkins University, Baltimore, MD, USA}
\email{mkarmen1@jhu.edu}

\author[0000-0001-5414-5131]{Mark Dickinson}
\affiliation{NSF NOIRLab, 950 N.\ Cherry Ave., Tucson, AZ 85719, USA}
\email{mark.dickinson@noirlab.edu}

\author[0000-0003-0212-2979]{Volker Bromm}
\affiliation{Department of Astronomy, The University of Texas at Austin, Austin, TX, USA}
\affiliation{Cosmic Frontier Center, The University of Texas at Austin, Austin, TX, USA}
\affiliation{Weinberg Institute for Theoretical Physics, University of Texas at Austin, Austin, TX 78712, USA}
\email{vbromm@astro.as.utexas.edu}

\author[0000-0001-8534-7502]{Bren E. Backhaus}
\affil{Department of Physics and Astronomy, University of Kansas, Lawrence, KS 66045, USA}
\email{bren.backhaus@ku.edu}

\author[0000-0003-1564-3802]{Marco Chiaberge}
\affiliation{Space Telescope Science Institute for the European Space Agency (ESA), ESA Office, 3700 San Martin Drive, Baltimore, MD, USA }
\affiliation{The William H. Miller III Department of Physics \& Astronomy, Johns Hopkins University, Baltimore, MD, USA }
\email{marcoc@stsci.edu}

\author[0000-0003-3881-1397]{Olivia R. Cooper}\altaffiliation{NSF Astronomy and Astrophysics Postdoctoral Fellow}
\affiliation{Department for Astrophysical and Planetary Science, University of Colorado, Boulder, CO 80309, USA}
\email{olivia.cooper@colorado.edu}

\author[0009-0007-8764-9062]{Yukta Ajay}
\affiliation{The William H. Miller III Department of Physics \& Astronomy, Johns Hopkins University, Baltimore, MD, USA }
\email{yajay1@jh.edu}

\author[0000-0001-6813-875X]{Guillermo Barro}
\affiliation{University of the Pacific, Stockton, CA 90340 USA}
\email{gbarro@pacific.edu}

\author[0000-0002-4153-053X]{Danielle A.\ Berg}
\affiliation{Department of Astronomy, The University of Texas at Austin, Austin, TX 78712}
\affiliation{Cosmic Frontier Center, The University of Texas at Austin, Austin, TX 78712}
\email{daberg@austin.utexas.edu}

\author[orcid=0000-0003-1051-6564]{Jenna Cann}
\affiliation{X-ray Astrophysics Laboratory, NASA Goddard Space Flight Center, Code 662, Greenbelt, MD 20771, USA}
\affiliation{Center for Space Science and Technology, University of Maryland, Baltimore County, 1000 Hilltop Circle, Baltimore, MD 21250, USA}
\affiliation{Center for Research and Exploration in Space Science and Technology, NASA/GSFC, Greenbelt, MD 20771}
\email{jenna.cann@nasa.gov}

\author[0000-0003-1371-6019]{M. C. Cooper}
\affiliation{Department of Physics \& Astronomy, University of California, Irvine, 4129 Reines Hall, Irvine, CA 92697, USA}
\email{cooper@uci.edu}

\author[0000-0001-9440-8872]{Norman A. Grogin}
\affiliation{Space Telescope Science Institute, 3700 San Martin Drive, Baltimore, MD 21218, USA}
\email{nagrogin@stsci.edu}

\author[0000-0002-3301-3321]{Michaela Hirschmann}
\affiliation{Institute of Physics, Laboratory for galaxy evolution, EPFL, Observatory of Sauverny, Chemin Pegasi 51, 1290 Versoix, Switzerland}
\email{michaela.hirschmann@epfl.ch}

\author[0000-0002-1416-8483]{Marc Huertas-Company}
\affiliation{Instituto de Astrofísica de Canarias, c/ Vía Láctea sn, 38025 La Laguna, Spain} \affiliation{Universidad de La Laguna. Avda. Astrofísico Fco. Sanchez, La Laguna, Tenerife,
Spain}
\email{mhuertas@iac.es}

\author[0000-0001-9187-3605]{Jeyhan S. Kartaltepe}
\affiliation{Laboratory for Multiwavelength Astrophysics, School of Physics and Astronomy, Rochester Institute of Technology, 84 Lomb Memorial Drive, Rochester, NY 14623, USA}
\email{p@nasa.gov} 

\author[0000-0002-6610-2048]{Anton M. Koekemoer}
\affiliation{Space Telescope Science Institute, 3700 San Martin Drive, Baltimore, MD 21218, USA}
\email{koekemoer@stsci.edu}

\author[0000-0003-1581-7825]{Ray A. Lucas}
\affiliation{Space Telescope Science Institute, 3700 San Martin Drive, Baltimore, MD 21218, USA}
\email{lucas@stsci.edu}

\author[0000-0002-7530-8857]{Arianna S. Long}
\affiliation{Department of Astronomy, The University of Washington, Seattle, WA USA}
\email{}

\author[0000-0002-5222-5717]{Roberto Gilli}
\affiliation{INAF$-$ Osservatorio di Astrofisica e Scienza dello Spazio di Bologna, Via P. Gobetti 93/3, 40129 Bologna, Italy}
\email{}

\author[0000-0002-5222-5717]{Colin Norman}
\affiliation{Space Telescope Science Institute, 3700 San Martin Drive Baltimore, MD 21218, USA}
\affiliation{Department of Physics \& Astronomy, Johns Hopkins University, Bloomberg Center, 3400 N. Charles St., Baltimore, MD 21218, USA}
\email{}

\author[0000-0001-5655-1440]{Andrew F. Ptak}
\affiliation{NASA-Goddard Space Flight Center, Code 662, Greenbelt, MD, 20771, USA}
\email{}

\author[0000-0002-3703-0719]{Chris T. Richardson}
\affiliation{Elon University,
100 Campus Drive,
Elon, NC 27244}
\email{crichardson17@elon.edu}

\author[0000-0000-0000-0001]{Jane R. Rigby}
\affiliation{Astrophysics Science Division, NASA Goddard Spaceflight Center, code 660, 8800 Greenbelt Rd, Greenbelt, MD 20771, USA}
\email{jane.rigby@nasa.gov}

\author[0000-0002-8163-0172]{Brittany N. Vanderhoof}
\affiliation{Space Telescope Science Institute, 3700 San Martin Drive, Baltimore, MD 21218, USA}
\email{bvanderhoof@stsci.edu}

\author[0000-0003-3466-035X]{{L. Y. Aaron} {Yung}}
\altaffiliation{Giacconi Postdoctoral Fellow}
\affiliation{Space Telescope Science Institute, 3700 San Martin Drive, Baltimore, MD 21218, USA}
\email{yung@stsci.edu}

\author[0000-0002-7051-1100]{Jorge A. Zavala}
\affiliation{University of Massachusetts Amherst, 710 North Pleasant Street, Amherst, MA 01003-9305, USA}
\email{jzavala@umass.edu}

\collaboration{all}{The THRILS and C3PO Collaborations}

\begin{abstract}

The enigmatic population of ``Little Red Dots'' (LRDs) is at the center of some of the largest debates in extragalactic astronomy today.  The source(s) of ionizing emission and the physical scale over which it governs is still largely unknown. We show for the first time spectroscopic variability in a $z \sim 7$ LRD. Comparing a recently obtained  \jwst/NIRSpec 10.2 hr F290LP/G395M spectrum via the C3PO survey to an 8.4 hr F290LP/G395M spectrum taken 99 days earlier ($\sim$13 rest-days) via the THRILS survey, we find a $\sim 30\% $ difference in the continuum and broad-line flux, and a $\sim 42\%$ difference between \oiiidbl\ flux in the two epochs. Through rigorous testing, we confirm that such differences are not the result of differing MSA slit placements on source nor merely flux calibration offsets. These results are further corroborated by both a similar continuum and distinct \oiiidbl\ flux differences found in NIRSpec prism/clear observations of the source at an epoch taken approximately a year earlier than the THRILS observations via RUBIES and an additional observation fortuitously taken during the THRILS epoch (within a rest-day) via the CAPERS survey. Assuming LRDs are a type of accreting black hole system, this implies direct sight-lines must exist from the accretion disk to the surrounding nebular gas on scales beyond the broad-line region, and thus any high-density gas interpretations must allow for covering fractions $< 100\%$. Furthermore, these results show the \oiii\ line emission is likely not galaxy process-dominated, with a significant population of the narrow-line emitting gas closest to the broad-line region being directly ionized by the LRD. Finally, these results highlight the need for new approaches in inferring black hole properties of these systems, accounting for the lack of significant ionization via star formation, and/or exploring more exotic host-galaxy conditions at these early epochs.

\end{abstract}



\section{Introduction} 

\defcitealias{deugenioirony}{D'Eugenio\,\,et\,\,al.\,\,2025b}
\defcitealias{Lambrides25}{Lambrides\,\,et\,\,al.\,\,2025}
\defcitealias{tang25}{Tang\,\,et\,\,al.\,\,2024}
\defcitealias{wang24evol}{Wang\,\,et\,\,al.\,\,2024}
\defcitealias{kocevski25}{Kocevski\,\,et\,\,al.\,\,2025}
\begin{table*}
\centering
\footnotesize
\begin{tblr}{
  width = \linewidth,
  colspec = {Q[25]Q[95]Q[73]Q[118]Q[105]Q[70]Q[104]Q[140]},
  hline{1,11} = {-}{0.08em},
  hline{2} = {-}{},
}
PID & PI & Dates & Instrument/Mode & Filter/Grating & $t_{exp}$ & Name Used & Reference\\
5943 & Papovich, Hu, Hutchison & 22-06-2025 & NIRSpec/MSA & F290LP/G395M & \textbf{10.2 hr} & GlimmIr & \textbf{this work}; Papovich et al.\ (in prep)\\
4106 & Nelson, Labbe & 22-04-2025 & NIRSpec/MSA & F290LP/G395M & 7.4 hr & Irony & \citetalias{deugenioirony}\\
6368 & Dickinson & 24-03-2025 & NIRSpec/MSA & clear/prism & 48.14 min & --- & ---\\
5507 & Hutchison, Larson & 16-03-2025 & NIRSpec/MSA & F290LP/G395M & \textbf{8.4 hr} & THRILS-46403, RUBIES\_49140 & \citetalias{Lambrides25}\\
4287 & Mason, Stark & 19-03-2024 & NIRSpec/MSA & F290LP/G395M & 58.35 min & CEERS-10444 & \citetalias{tang25}\\
4233 & de\,\,\,Graaff, Brammer & 13-03-2024 & NIRSpec/MSA & F290LP/G395M & 48 min & RUBIES-49140 & \citetalias{wang24evol}\\
4233 & de\,\,\,Graaff, Brammer & 13-03-2024 & NIRSpec/MSA & clear/prism & 48 min & RUBIES-49140 & \citetalias{wang24evol}\\
2279 & Naidu & 23-05-2023 & NIRCam/imaging & F444W & --- & --- & --- \\
1345 & Finkelstein & 28-06-2022 & NIRCam/imaging & F444W & --- & CEERS-10444 & \citetalias{kocevski25}
\end{tblr}
\caption{\label{table:observations}Table of \jwst\ NIRSpec and NIRCam observations, in reverse chronological order.}
\end{table*}

 Commonly assumed to be a class of accreting super-massive black holes (SMBHs), Little Red Dots (LRDs) are characterized by their extreme compactness, V-shaped spectral energy distribution whose nadir corresponds to the Balmer edge ($\sim$3645$\mathrm{\AA}$), broadened Balmer emission, and for a sub-set of sources, narrow Balmer absorption \citep{barro24,labbe25,matthee,harikane23, larson23,furtak,greene23,kokorev23, kocevski23,2024ApJ...968....4P,kocevski25,kokorev24, killi24,akins24,taylor25a,Lambrides25,wang24}. While the true nature of these sources remains contested, the ever-present broadened hydrogen emission indicates rapidly moving gas that is consistent with the broad-line region (BLR) that is within the gravitational influence of an accreting massive black hole (hereinafter active galactic nucleus or AGN). Each unique interpretation could herald a paradigm shift in our understanding of early black hole formation, growth, and subsequent evolution alongside their host galaxy \citep[e.g.,][]{Smith_rev2019,Inay20}. 

Attempts to converge on a unifying model of their true nature have led to several competing theories that can yield fundamentally different source properties, including black hole masses and bolometric luminosities \citep{deugenio25a,degraaff25,taylor25b,lambrides24b,inayoshi25,naidu25,ji2025blackthunder,2026arXiv260220247P, ronayne25,barro25}. Such diversity of theories makes it difficult to determine which, if any, scaling relations derived from more canonical samples of accreting SMBHs are applicable to these sources.

\begin{figure*}
    \centering
    \includegraphics[width=0.95\linewidth]{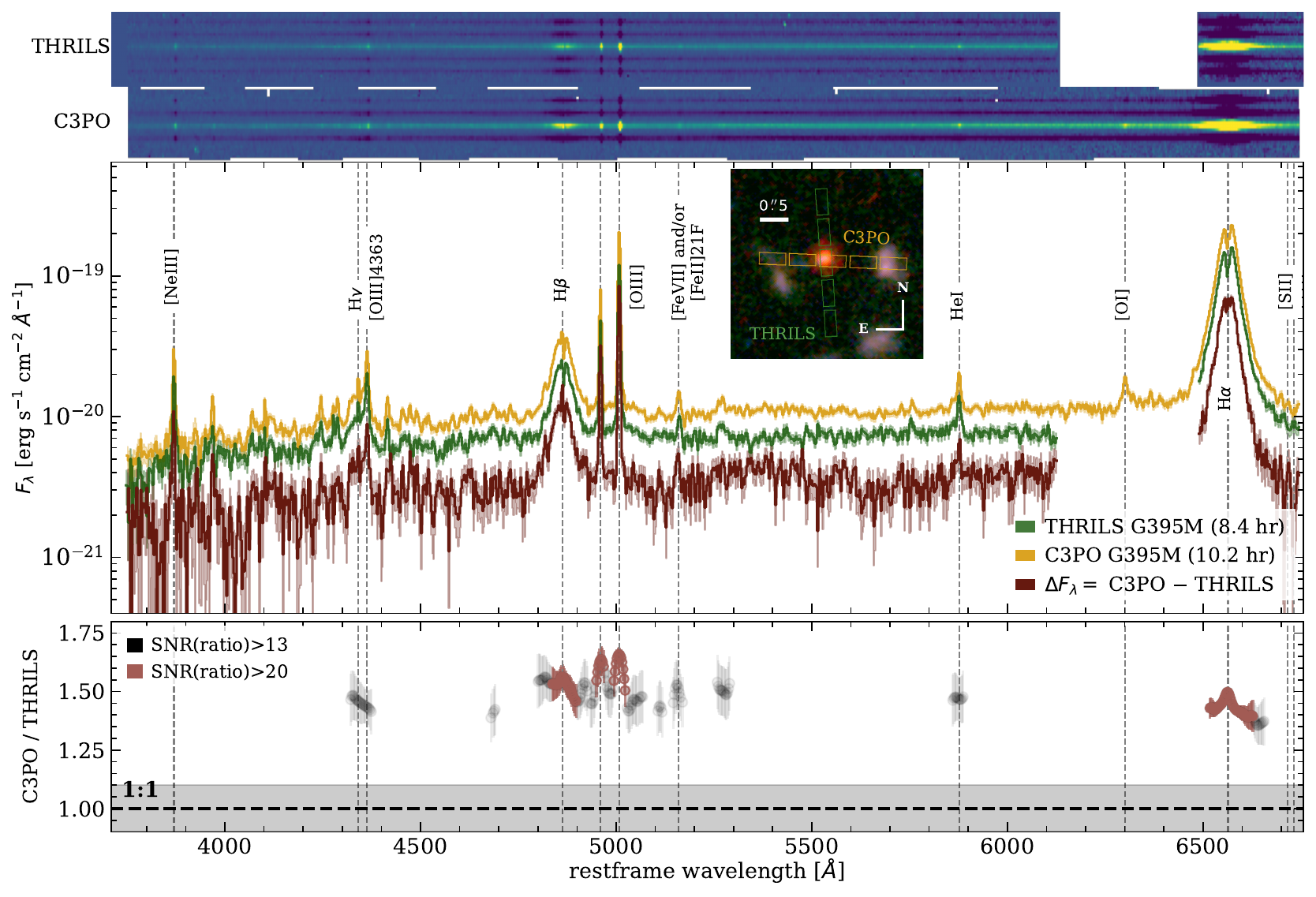}
    \caption{\textbf{Spectroscopic Variability in an Early Little Red Dot.} (\textit{top}) The C3PO and THRILS 2D G395M/F290LP spectra of the GlimmIr. (\textit{middle}) The deep G395M/F290LP 1D spectra, with the THRILS spectrum in green and the C3PO spectrum in gold. The subtraction between the two spectra is shown in red. In the inset, MSA slit configuration for both programs is overlaid on the CEERS-10444 image of the source.  As noted in Section 2.2, contaminated slitlets were discarded from the spectral extraction. (\textit{bottom}) The red spectrum represents the most significant per element ratios of the two epochs in the previous panel --  highlighting the factor $>1.5$ between C3PO and THRILS spectra in \halpha,\hbeta\ and the \oiiidbl.}
    \label{fig:bigfig}
\end{figure*}

AGN are found to vary in every wave band in which they have been studied on time scales from hours to decades \citep{vandenvar,macleod12}. Spectroscopic variability is a critical diagnostic tool that provides key insights on the structure and scale of the largely unresolved sub-pc regions close to the SMBH \citep{peterson1993,illic15,rashed15,sturm18}. For instance, reverberation mapping methods rely on characterizing the radius of the broad-line region by characterizing the time it takes for a continuum photon to ionize the broad-line region clouds, leading to the development of empirical power-law relationships between the radius of the BLR and the AGN continuum luminosity \citep{blandford82,peterson1993,shen24}. 

The narrow-line region is located on scales further than the BLR as is indicated by measured lower velocity dispersions in case of a virial velocity field and the existence of line species with critical densities much lower than the average BLR gas density \citep{gaskell09, zhu23}. Variability in the narrow-line region is much more difficult to observe, largely due its nature. The location of the inner radius of the narrow-line region is dependent on the luminosity of the AGN \citep{baskin05} and can begin on sub-pc scales and extend up to kpc scales \citep{peterson13}. In non-spatially-resolved spectroscopy (i.e., traditional ``slit'' spectroscopy), without the tell-tale kinematic signatures that the BLR contains, more sophisticated approaches are required to effectively isolate which components of observed emission lines are due to the narrow-line region of the AGN vs. host-galaxy processes like star-formation. Some studies have found evidence of weak and long narrow-line variability in broad-line sources \citep{peterson13,rashed15}, but in most multi-epoch surveys of AGN, spectroscopic variability is only observed in either the continuum, broadened emission components, and/or extremely high-ionization lines ($>$80 eV) that are predominantly associated with AGN ionization \citep{landt15,smith25}. 

Thus, as multiple JWST observation cycles have accrued, recent efforts have sought to determine whether LRDs exhibit characteristic variability \citep{zhou25,burke25,zhang25,venuslrdvar,lin26}, or should \citep{secunda26}. In terms of spectroscopic searches, only a handful of LRDs have been spectroscopically re-observed in similar instrument configurations to allow for such a systematic search. Recently, \cite{ji2025blackthunder,qso1var} inspected the spectroscopic variability of the multiply lensed LRD, A2744-QSO1. Lensing-induced time delays provided a unique opportunity to test for variability over a span of 22 yr in the rest-frame. In particular, \citet{qso1var} found that the equivalent widths of both the \halpha\ and \hbeta\ broad components exhibited $\sim$20\% flux variations over the course of the observations.

A caveat in conclusively interpreting the nature of the flux differences in A2744-QSO1 is the fact that only two emission lines were reported as significantly detected, thus it is challenging to determine if the flux differences are driven by continuum or line change. The over-arching conclusion from A2744-QSO1 \citep{qso1var}, and photometric variability searches \citep{venuslrdvar}, is that if variability is present it occurs on decade time-scales. Unfortunately, due to the lack of spectroscopic re-observation of bright LRDs in similar instrument set-ups and at sufficient depths make it difficult to comment on the full diversity of spectroscopic variations potentially present in these sources.

In this letter, we present for the first time, spectroscopic variability in both the broad- and narrow-line components of this LRD. We detail the numerous checks performed to test if any systematic effects are driving the flux differences between spectra, and in the discussion we detail the implications of these results to the larger LRD landscape in the early cosmos.

\begin{figure*}
    \centering
    \includegraphics[width=0.95\linewidth]{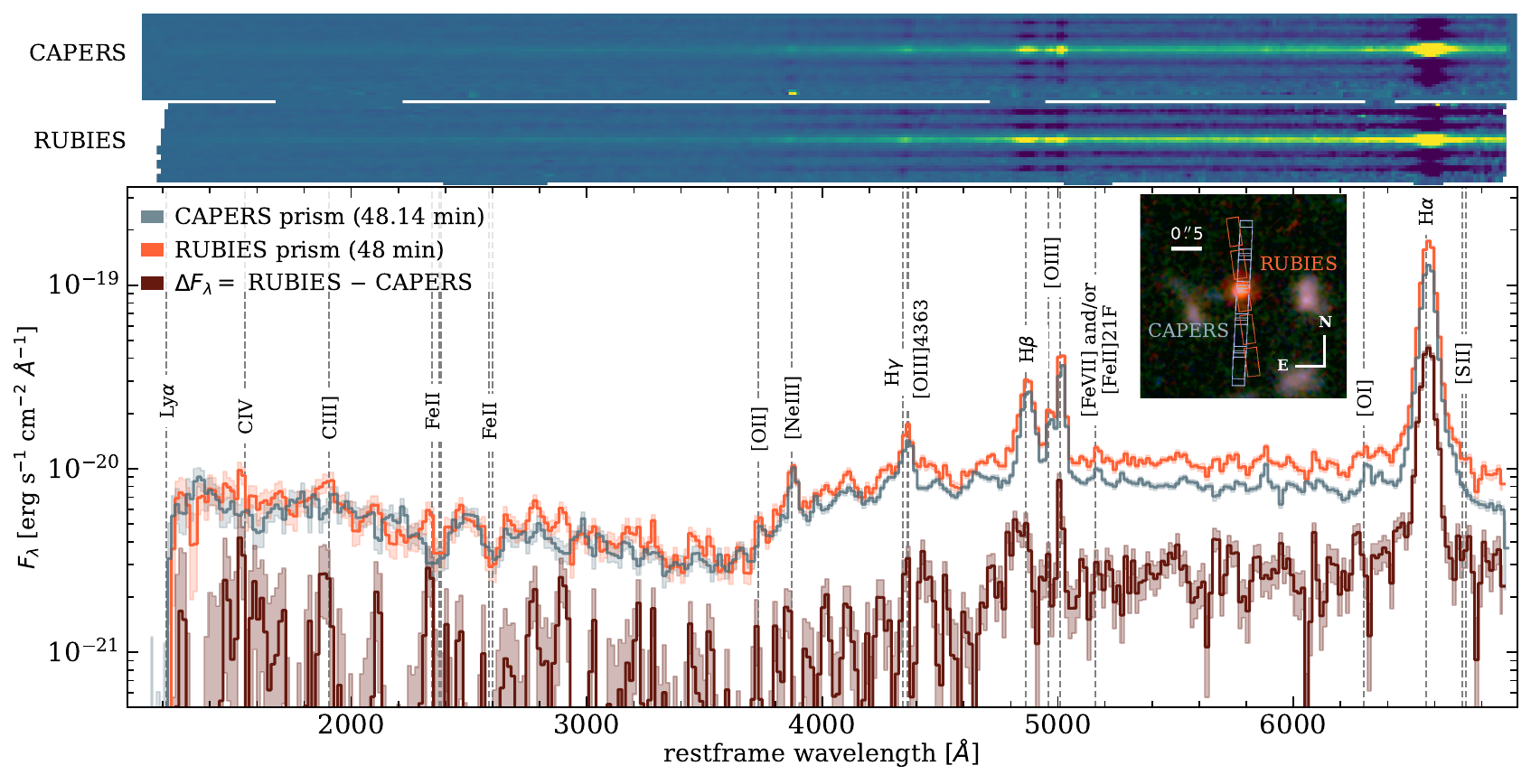}
    \caption{\textbf{Spectroscopic Variability in an Early Little Red Dot.} (\textit{top}) The CAPERS and RUBIES 2D prism/clear spectra of the GlimmIr. (\textit{bottom}) The CAPERS spectrum is in blue and the RUBIES spectrum is in orange. The subtraction between the two spectra is shown in red. In the inset, MSA slit configuration for both programs is overlaid on the CEERS-10444 image of the source.  As noted in Section 2.2, contaminated slitlets were discarded from the spectral extraction}
    \label{fig:bigfig-prism}
\end{figure*}

\section{Multi-Epoch Deep G395M Observations of the GlimmIr} \label{sec:data}

We refer to the source that is the subject of this work as GlimmIr -- which is a combination of both Glimmer and \textit{Irony}, a recent naming from \citet{deugenioirony}. Due to the rich set of observations of the GlimmIr, subsequent re-identifications, and to fully account for the source's provenance, we provide the program information and associated publications in Table \ref{table:observations}. To summarize, this source was initially photometrically identified in CEERS (ERS-1345, PI: Finkelstein), and later initially classified as an LRD with ID CEERS-10444 in \cite{kocevski25}. We note its spectral confirmation as an LRD was made possible by RUBIES (GO-4233, PIs: A. de Graaff \& G. Brammer; \citealt{degraaff25}) where it was spectroscopically targeted for the first time. The RUBIES observations comprised both  \textit{JWST}/NIRSpec multi-shutter array \citep[MSA;][]{ferruit22} low-resolution prism (R$\sim$100) and medium-resolution (R$\sim$1000) F290LP/G395M gratings, and was identified by its ID as RUBIES\_49140 \citep{wang24evol} and subsequently observed in GO 4287 (PIs: C. Mason \& D. Stark). Ionization line analysis of this source was presented in \cite{tang25, wang25}. The deepest medium resolution observations of the source were acquired as part of the THRILS (GO-5507, PIs T.\ Hutchison \& R.\ Larson; \citealt{hutchison25b}) and C3PO (GO-5943, PIs C.\ Papovich, W.\ Hu, \& T.\ Hutchison) programs with 8.4 hr and 10.2 hr {\it JWST}/NIRSpec spectroscopy, respectively, in the G395M medium resolution (R$\sim$1000) grating. Additional NIRSpec prism observations were conducted on the source in CAPERS (GO-6368; PI: M. Dickinson). The THRILS spectrum of the GlimmIr was recently published under the dual identification of the MSA ID in THRILS and RUBIES (THRILS-46403/RUBIES-49140) in \citet{Lambrides25}. In between the THRILS and C3PO observations, the source was also re-observed as a part of a failed observation within GO 4106 (PIs: E.\ Nelson \& I. Labbe) with a 7.4 hr G395M observation and presented as \textit{Irony} in \citep{deugenioirony}.

\begin{figure*}[t]
    \centering
    \minipage{0.705\textwidth}
    \includegraphics[trim=4.5 0 0 0,clip,width=\linewidth]{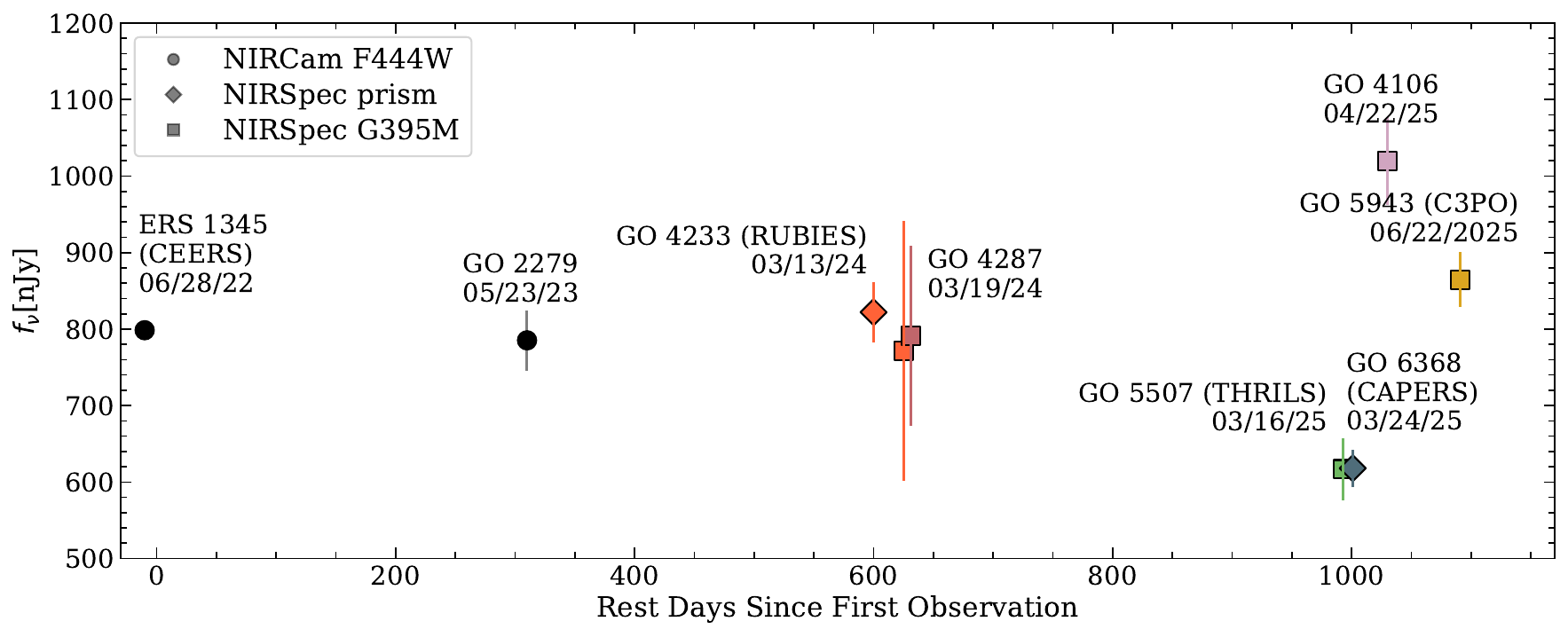}
    \endminipage\hfill
    \minipage{0.29\textwidth}
    \includegraphics[trim=13 3 0 0,clip,width=\linewidth]{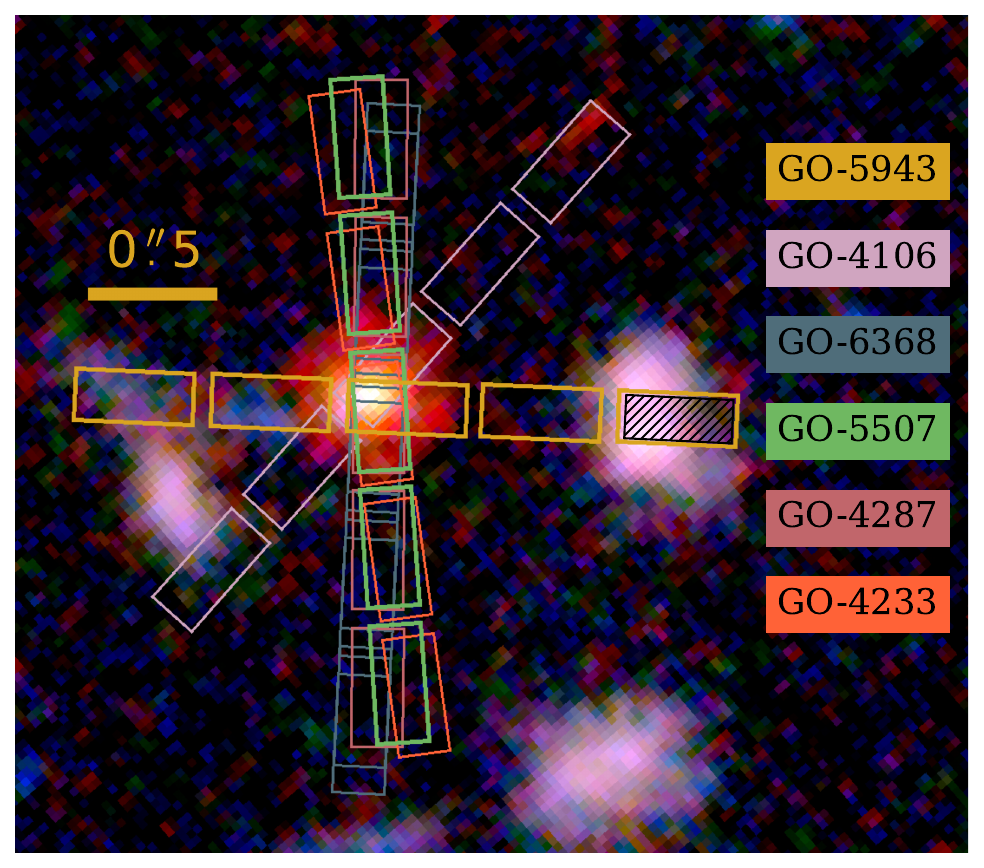}
    \vspace{1mm}
    \endminipage
    \caption{(\textit{left})Synthetic F444W for all NIRSpec Observations: Circles correspond to NIRCam imaging, diamonds correspond to prism observations, squares correspond to G395M observations. Colors of non-NIRCam points correspond to the slit orientations shown in the right-most panel. (\textit{right}) Slit Orientations for all multi-epoch observations of the GlimmIR (see Table \ref{table:observations})}
    \label{fig:placeholder}
\end{figure*}

In order to robustly compare the observations of our source of interest from all the various programs at different epochs, we carry out a complete homogeneous reduction of the source in all of the available observations (see Table \ref{table:observations}). Note that the data from GO 4106 is still not public, so we use the spectrum of the source published in \citet{deugenioirony}. 
The reduction is performed making use of the \texttt{jwst} reduction pipeline\footnote{\url{https://github.com/spacetelescope/jwst}} version 1.20.2 \citep{BushousePipeline1.20.2} with reference files from the \texttt{jwst\_1464.pmap} CRDS context.
We follow the general methodology described in \citet{ArrabalHaro23}, with a few additional steps for specific improvements as summarized below. 
The \texttt{clean\_flicker\_noise} function of the \texttt{Detector1Pipeline} module is employed for an improved treatment of the 1/$f$ noise at the detector-level, using the NSClean algorithm for the fit method \citep{nsclean}. 
We make use of a modified \texttt{FFLAT} reference file for the NIRSpec fore-optics flat field to avoid propagating the unrealistically large errors currently present at certain wavelengths in the default \texttt{FFLAT} files.
An exhaustive description of this data reduction workflow will be presented in a future work.
Shutters that could have been closed by any kind of unexpected failure during the MSA observations of any of the programs are identified based on the lack of sky emission in background-unsubtracted versions of the 2D spectra. We find that the source of interest was not affected by any unexpected closed shutter in any of the programs it was observed.
Moreover, a tailored nodded background subtraction pattern is employed for each observation (at different aperture angles) in order to carefully avoid any kind of possible over-subtraction from neighbor objects entering any of the shutters in some of the nods.
The final 1D spectra are optimally extracted from the rectified 2D spectra following the prescription in \citet{Horne86}. Additionally, to further study possible systematic variability in the flux calibration, the object was reduced both enabling and disabling the \texttt{PATHLOSS} correction in the \texttt{Spec2Pipeline} module, which accounts for the spatially dependent flux calibration for a point-like source as a function of its location within the shutter hosting it. We find the difference between the corrected and uncorrected flux to be smaller ($<10\%$) than the variability measurements detailed in Section \ref{sec:vari}. The associated figure of this test is presented in Appendix \ref{fig:pathloss}.

\subsection{Spectral Line Fitting}
Before fitting spectral features, we first linearly re-bin both spectra to the same wavelength grid while conserving flux. We then fit the spectra using version 3 of the open-source Python 3 code Bayesian AGN Decomposition Analysis for Spectra (BADASS; \citealt{sexton21}). In brief, BADASS implements emcee \citep{emcee} to obtain robust parameter fits and parameter uncertainties, then utilizes a custom autocorrelation analysis to assess parameter convergence. Each spectral fit was run for a maximum of 25,000 MCMC iterations, with the mean of parameters converging around 20,000 iterations. 

For the THRILS and C3PO, we fit for absorption as well as broadened and narrow emission for all Balmer complexes detected in both epochs i.e., \hgamma, \hbeta, and \halpha. In \hgamma, we also fit for the auroral \oiii\ \lam4364 line and \feii\ 21F 4359; in \hbeta, \feii\ 20F 4816; and in \halpha\ the two \nii\ lines. The final decomposition of all Balmer lines is well converged, and the spread of the posteriors for the dispersion, amplitude, and velocity offset of all parameters is within 10\% of their respective values. We anchor the velocity offsets of all non iron emission lines to the \oiii, as well as anchoring all the velocity offsets of \feii\ to each other. In addition to these lines, we also fit every line previously detected at $>$5$\sigma$ in \citet{Lambrides25} for this source. 

\begin{figure*}
    \centering
    \includegraphics[width=\linewidth]{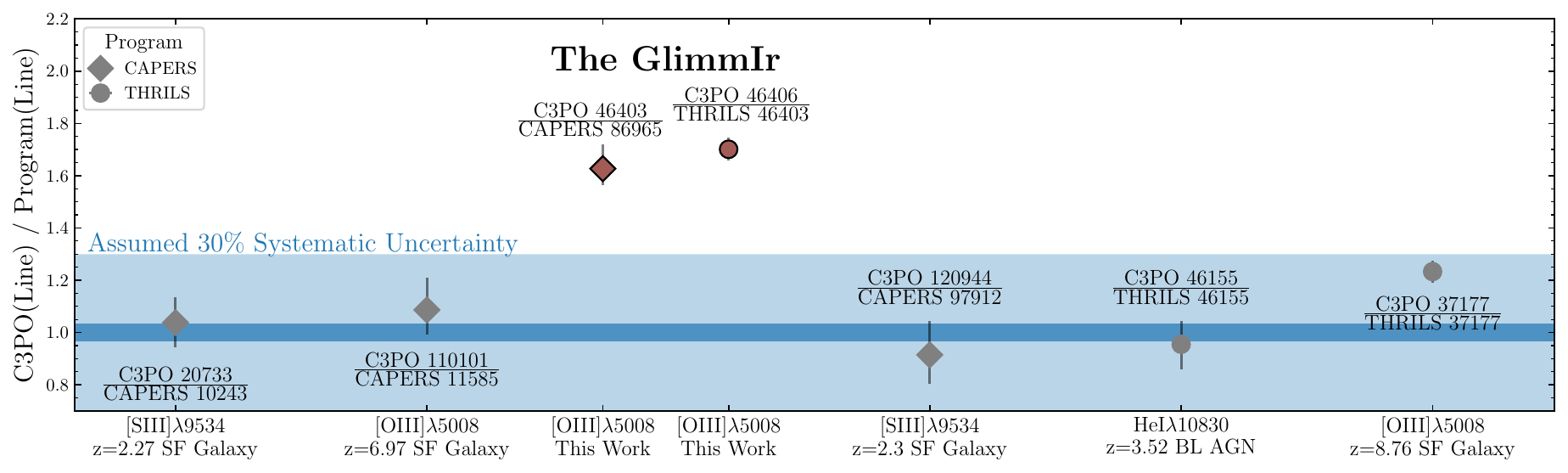}
    %
    \caption{\textbf{Comparing to other Dually Observed Sources}: Flux ratio of brightest nebular line for sources that are similarly centered to both THRILS+C3PO GlimmIr observations and CAPERS+C3PO GlimmIr observations. Blue shaded area is an assumed 30\% systematic flux difference. The GlimmIr, particularly the C3PO/THRILS epoch, is a $> 5 \sigma$ outlier compared to all other sources. We note the CAPERS and THRILS epochs were observed within a rest day of each other. }
    \label{fig:allspec}
\end{figure*}

Due to the increase in observation depth of 2 hr between C3PO and THRILS, several marginally detected lines in the THRILS spectrum were robustly detected in the C3PO. The full line analysis of this spectrum will be presented in an upcoming work (Lambrides+26, in prep). In this work, we conservatively only compare the flux differences between the highest-significance features ($>{5}\sigma$ ) detected in both epochs of observations and the sources with the most well-constrained fluxes from the model fitting $>{5}\sigma$. This latter distinction is especially relevant for the Balmer complexes. For instance, while the total flux in the \hgamma\ region ($4300\mathrm{\AA} - 4400\mathrm{\AA}$) is well detected, the line fit requires four individual gaussian components due to the blending of the Balmer emission with \oiiiauroral\ and \feii. While the statistical certainty of the individual line fit may be well constrained enough to recover the total flux of the complex, the degeneracy due to line blending increases the systematic uncertainty. These uncertainties propagate when comparing the line flux differences between the model fits for each epochs, and thus we constrain our variability analysis, unless otherwise stated, to the lines and/or features whose modeled flux uncertainty is $>{5}\sigma$.    

%

\section{Variability in the GlimmIr} \label{sec:vari}
\begin{deluxetable}{lccc}[h!]
\tablecolumns{4}
\tablecaption{\label{table:linefluxes}Table of Compared Line Fluxes from THRILS \& C3PO }
\tablehead{ \colhead{Line} & \colhead{THRILS} & \colhead{C3PO} & \colhead{C3PO/THRILS} }
\startdata
     & & & \\
L$_{\mathrm{cont},5100}$ 10$^{43}$[erg/s] & 1.98$^{+0.03}_{-0.03}$ & 2.87$^{+0.03}_{-0.02}$ & 1.45$^{+0.03}_{-0.02}$ \\
\halpha\ (Broad) & 723$^{+9}_{-8}$ & 1083$^{+7}_{-10}$ & 1.50$^{+0.02}_{-0.02}$ \\
\hbeta\ (Broad) & 67.66$^{+1.18}_{-1.14}$ & 108.2$^{+1.5}_{-1.5}$ & 1.60$^{+0.04}_{-0.03}$ \\
\oiii\ 5008 & 54.8$^{+0.4}_{-0.4}$ & 93.4$^{+0.5}_{-0.7}$ & 1.70$^{+0.02}_{-0.02}$ \\
\oiii\ 4960 & 18.20$^{+0.16}_{-0.19}$ & 31.0$^{+0.2}_{-0.2}$ & 1.70$^{+0.02}_{-0.02}$ \\
 & & & \\
\enddata
\tablecomments{ Line fluxes are reported with normalization $10^{-19}$ erg s$^{-1}$ cm$^{-2}$ {\AA}$^{-1}$. All errors are the 68\% confidence intervals. }
\end{deluxetable}

\subsection{Multi-epoch photometry}

While no photometry exists during the epochs where the majority of the spectra were taken, we can test if this source shows any variability before the spectra were taken. Two F444W images were taken of this source, the initial CEERS imaging (120 rest-days before the THRILS observation) and a direct image from a slitless-spec campaign taken 40 rest-days later (GO 2279). We re-reduce the CEERS pointing and GO 2279, and re-calculate the photometry in each epoch self-consistently using the methods outlined in \citet{ceerssteve}. We find no statistically significant difference between the first two photometric epochs ($< 3\sigma$). This either implies a) we caught a flare that occurred after GO 2279 b) the continuum varies on shorter timescales than the time separation between CEERS and GO 2279, thus catching it during a maximal difference in flux is more statistically unlikely, and/or c) we simply do not have sufficient cadence of observations to robustly quantify the nature of the variability. Thus, further multi-epoch spectro-photometric follow-up is critical for this source. 

As seen in Figure 1, we find significant differences between the THRILS and C3PO epoch of observations. The C3PO spectrum (gold) was taken 99 observed days (or 12.89 rest light-days) after the THRILS observation (green). The subtracted flux, shown in the panel below (red) already highlights the difference between the two spectra is not just a simple scaling of the flux as seen in the \halpha\ zoom-in shown in panel 2.

Each Balmer complex (\hgamma, \hbeta, and \halpha) in total exhibits a $\sim$30\% flux difference in the broad components of their respective Balmer complexes. The 2 hr increase in integration time also affords for the additional detection of \feii\ features in both the \hgamma\ and \hbeta\ line complexes. Interestingly, we broadly find the profile shape and line centers of the absorption components to be roughly the same, with only a marginal increase in the depth of the absorption line flux difference in the \hbeta\ and \halpha\ features. For instance, when we compare the difference between the peak flux of the \halpha\ emission and the peak flux of the measured \halpha\ absorption in both C3PO and THRILS, we find the ratio of the differences ($\sim22\%$) is below the assumed 30\% flux  uncertainty.  

Most surprisingly,we find at least \textit{a factor of 1.7} difference in both \oiii\ \lam5008 and \neiii, as seen in Table \ref{table:linefluxes}. We discuss in Section 4 the implications of these measurements, which provide unique constraints on the nature of the high density gas responsible for the Balmer self-absorption. The same level of flux differences found for \halpha, our most constrained broad line feature, is within the error of the continuum variability increase. 

We then test if the flux difference is truly intrinsic or due to unaccounted for systematic differences between the two different observing programs and thus different position angles of the telescope. If we are simply seeing differences in the flux due to the location on the slit, then an independent observation within one of the epochs should also have a flux difference. CAPERS was taken eight days (1.04 rest days) after the THRILS observations (see Table \ref{table:observations}). CAPERS, while shallower data (~5 hrs) and with lower resolution, (PRISM, R$\sim$100), detects the \oiiidbl\ at $>10$ SNR. We find the \oiiidbl\ is statistically similar between CAPERS and THRILS. Furthermore, we can also compare the fully scatter of flux differences between other sources that were also observed within either CAPERS--C3PO or THRILS--C3PO. 

To make the most direct comparison, we choose sources with similar offsets and locations on the slit to the GlimmIr in each respective observation that also have a $> 5$ SNR line in each observation. As shown in Figure \ref{fig:allspec}, this yields 3 additional sources observed in both CAPERS + C3PO and two additional sources observed in both THRILS+C3PO.  Even when using the derived 30\% systematic error on the flux ratio \citep[see][for deeper discussion on slit-loss]{dalmasso25}, the GlimmIr remains a $>$ 5 sigma outlier towards all other similar multi-epoch spectroscopic observations.

\section{Summary and Discussion}

\begin{figure}[t]
    \centering
    \includegraphics[width=\linewidth,trim=20 0 400 0, clip]{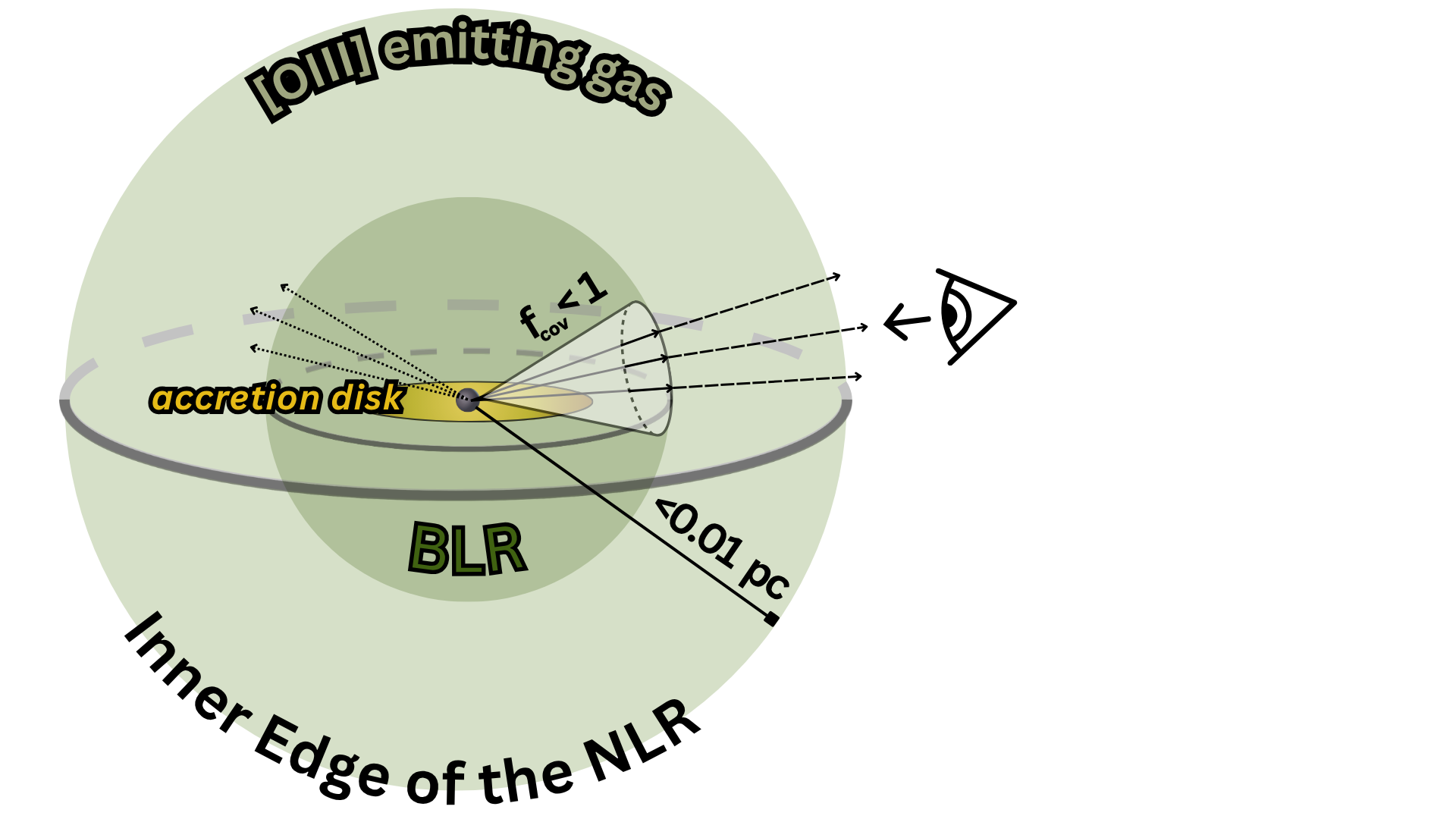}
    \caption{The existence of direct sight-lines from the accretion indicates a covering fraction $< 1$ of the broad-line gas. }
    \label{fig:schematic}
\end{figure}

In summary, we find a statistically significant flux increase of at least $>1.65$ for all strongly detected emission lines, and note a $>10 \sigma$ differences between the broad \halpha\ flux and \oiiidbl\ between the epochs. These results imply several important constraints on the nature of this source: 

\begin{enumerate}
    \item \textbf{A Direct Constraint on the Narrow-Line Region}: With only two epochs, it is difficult to conclusively infer the exact geometry of this source, but the fact we see any variability in strongly detected narrow lines like \oiiidbl\ provides significant insight. The THRILS and C3PO epochs are observed 99 days apart, which would correspond to rest 12.89 light days or $\sim 3.3 \times 10^{16}$ cm (0.01 pc). If one were to assume that the nebular lines like \oiii\ \lam5008 are within a population of gas that is in local thermal equilibrium (LTE), we are in density regimes well below the canonical broad-line region (i.e $n_{\mathrm{crit},[\mathrm{OIII}]5007} \sim 10^{5}~{\rm cm}^{-3}$) at or below 0.01 pc. In the event we are in non-LTE and/or the narrow line variability is driven by rapid changes to the local physical state of the gas (i.e density or opacity variations associated with clumping, instabilities, or bulk flows), these results imply inferring the cloud properties near the BLR using 
    standard prescriptions and/or values are not applicable. Future work involving higher cadence of observations and further radiative transfer simulations are required to robustly link the narrow line flux differences measured here to their physical implications on the size and ionization of the emitting region. 
    
    \item \textbf{Stellar processes are not driving the strong narrow line emission due to \oiii\ line variability:} We see variability in the \oiiidbl\ emission in lockstep with the broad lines -- thus evidence the \oiii\ emission is tied to the central engine of the LRD. Prior to these results, the nature of the narrow emission lines in LRDs underwent debate. In canonical AGN, lines with ionization potentials $<50$ eV, can be a mix of AGN and non-AGN processes alike. For lines like \oiii, it is has been largely found that AGN processes dominate \citep{laoroiii}. With LRDs, recent works suggest that \oiii\ may instead be driven by host galaxy processes. This was determined by comparing the Balmer break strength, or the the strength of the "v-shape" characterized in LRDs to the equivalent width (EW) of the \oiii \lam5008 line. In several recent works, an anti-correlation was found which was the opposite from the Balmer break strength vs \halpha\ EW \citep{degraff_oiii,barro25}. Our results indicate, that at least for the GlimmIr, the \oiii\ luminosity is directly related to accretion disk processes. 

    \item \textbf{The Need For Direct Sight-Lines:} In \citet{Lambrides25}, the detection of several transitions of \feii, in addition to a potential \fevii\ line, indicated that there must be unobstructed sight-lines to the accretion disk to excite these features. In particular, the \feii\ multiplets that were observed are most commonly found in AGN systems, and the lack of evidence of significant stellar evolution or other permitted transitions of \feiipermitted\ heavily discouraged a stellar origin for their excitation. This was in direct contention with LRD interpretations that require the covering fraction of ultra-dense gas ($>$ 10$^{11}$ cm$^{-3}$) to be near unity. Additionally, the flux differences to the broad and narrow emission lines of this source indicate a persistent sight-line on the order of at least 13 days must exist. As shown in Figure \ref{fig:schematic}, our current results imply that the covering fraction of the dense gas responsible for the Balmer absorption must be less than one for at least a sub-set of LRDs.  
\end{enumerate}

\begin{acknowledgments}
We thank Rachel Bezanson, Joel Leja, and Ivo Labbe for excellent discussions and key thoughts regarding the results in this work. This work uses observations made with the NASA/ESA/CSA \emph{JWST}. The data were obtained from the Mikulski Archive for Space Telescopes at the Space Telescope Science Institute, which is operated by the Association of Universities for Research in Astronomy, Inc., under NASA contract NAS 5-03127 for \jwst. These observations are associated with \jwst\ Cycle 3 GO programs 5507 and 5943. Support for the programs JWST-GO-5507 and JWST-GO-5943 was provided by NASA through a grant from the Space Telescope Science Institute, which is operated by the Association of Universities for Research in Astronomy (AURA), Incorporated, under NASA contract NAS5-26555.
The material is based upon work supported by NASA under award number 80GSFC24M0006.

ELL's and TAH's research is supported by an appointment to the NASA Postdoctoral Program at the NASA Goddard Space, administered by Oak Ridge Associated Universities under contract with NASA, as well as the University of Maryland Baltimore County and the Center for Space Sciences and Technology.
R.L.L and LYAY appreciates support from a Giacconi Fellowship at the Space Telescope Science Institute, which is operated by the Association of Universities for Research in Astronomy, Inc., under NASA contracts NAS 5-26555 and NAS5-03127.
PAH acknowledges support from NASA under award 80GSFC24M0006. The Center for Computational Astrophysics at the Flatiron Institute is supported by the Simons Foundation. 
SEIB is supported by the Deutsche Forschungsgemeinschaft (DFG) under Emmy Noether grant number BO 5771/1-1. P.G.P.-G. acknowledges support from grant PID2022-139567NB-I00 funded by Spanish Ministerio de Ciencia e Innovaci\'on MCIN/AEI/10.13039/501100011033, FEDER {\it Una manera de hacer Europa}.
\end{acknowledgments}

\clearpage
\appendix

\section{Pathloss Impact Comparison}

\begin{figure}[h]
    \centering
    \includegraphics[width=0.95\linewidth]{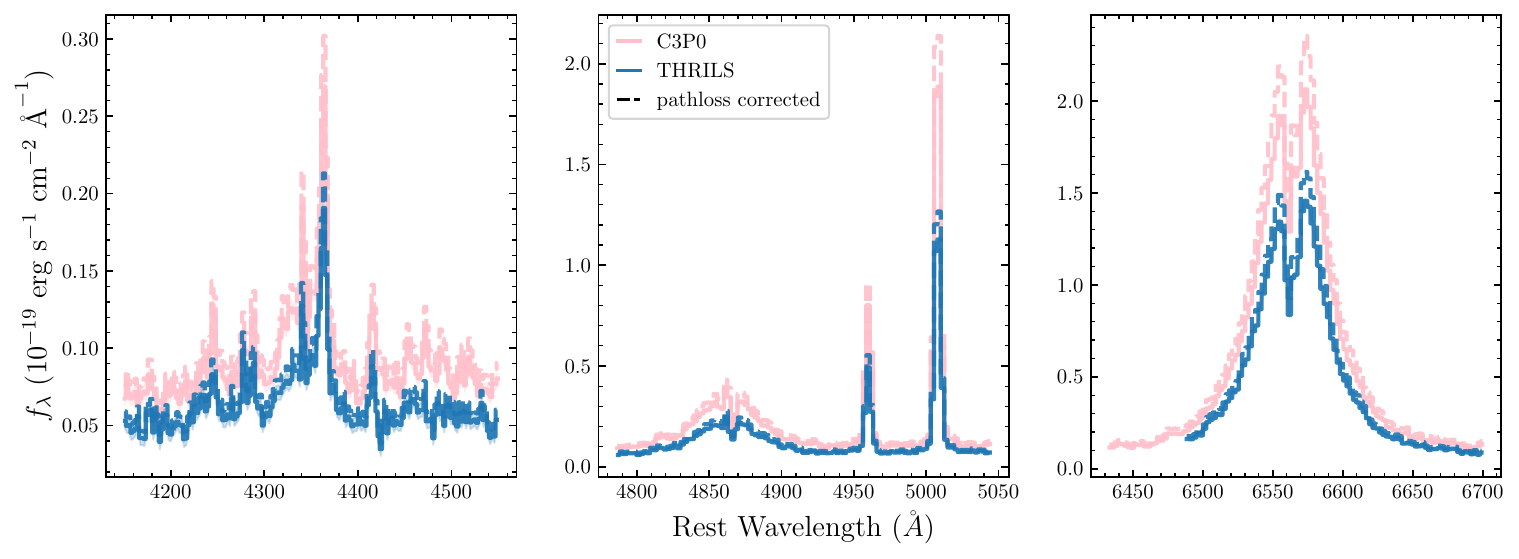}
    \caption{Pathloss Impact: The differences in flux between the three spectral regions shown are greater than the differences in flux due to including or not including a pathloss correction.}
    \label{fig:pathloss}
\end{figure}

As described in Section \ref{sec:data} and shown in Figure \ref{fig:pathloss}, we find the difference between the pathloss corrected and uncorrected flux to be smaller ($<10\%$) than the variability measurements detailed in Section \ref{sec:vari}. 

\section{Extraction Size Test}

\begin{figure}[h]
    \centering
    \includegraphics[width=.5\linewidth]{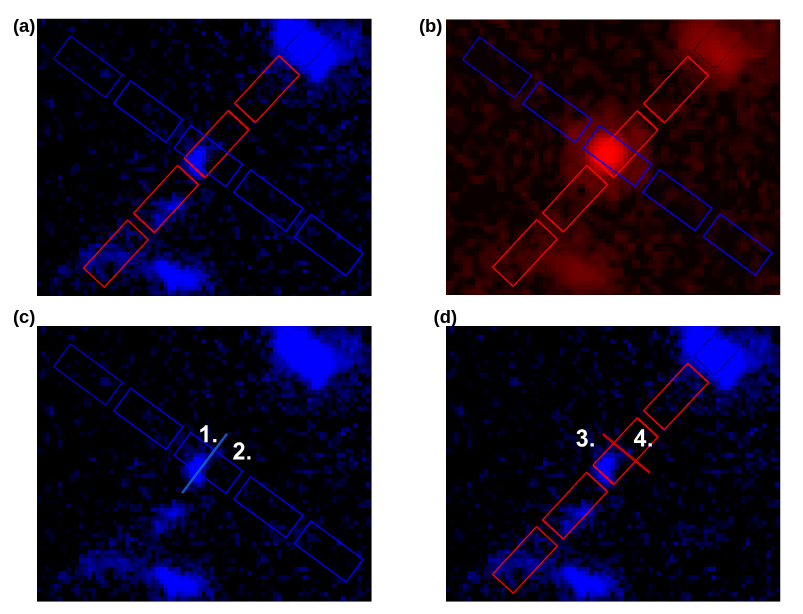}
    \caption{(a) F115 (blue slitlets are THRILS and red slitlets are C3PO) (b) F444W overlapped -- note the center of the F444W emission is offset from the center of the F115W blob. (c) The two sub-aperture definitions for THRILS, and (d) the two sub-aperture definitions for C3PO.}
    \label{fig:subap}
\end{figure}

\begin{figure}
    \centering
    \includegraphics[width=1\linewidth]{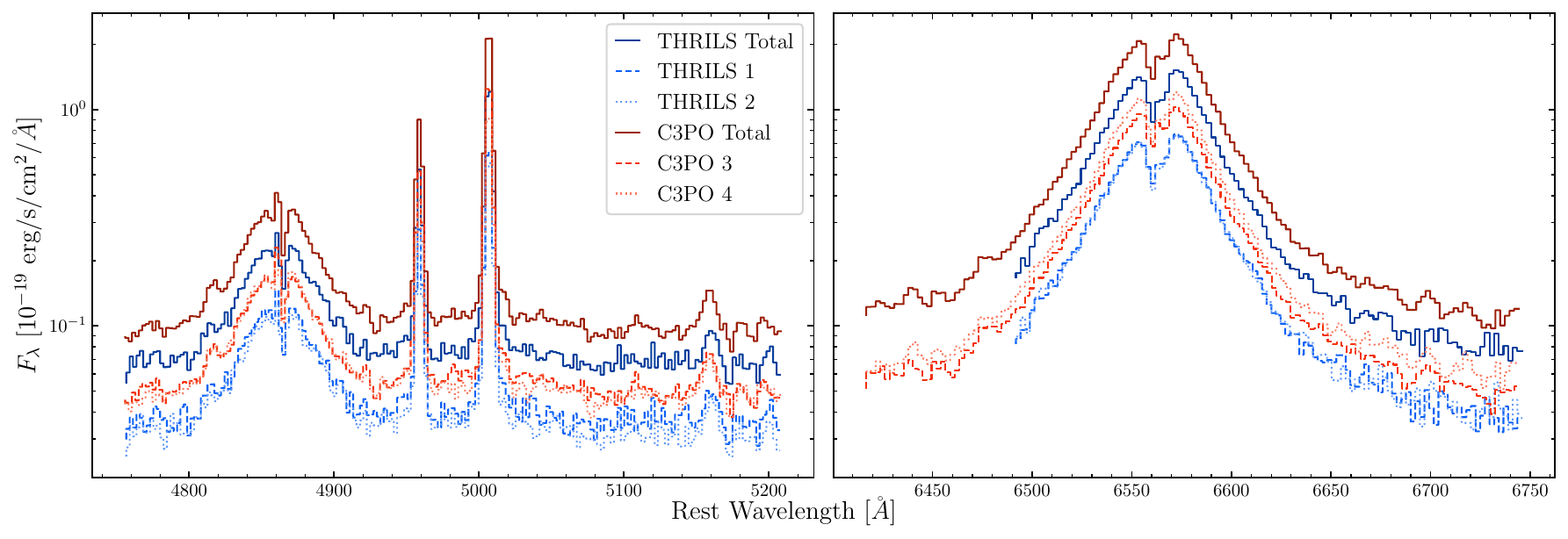}
    \caption{Spectra labels are defined in Section \ref{fig:subap}. C3PO(4) is slightly greater than C3PO(3) -- due to the brightness of \halpha, we are potentially seeing the impact of sub-sampling on (4) or off (3) center of the AGN. }
    \label{fig:subap2}
\end{figure}

There are some known resolved blue features -- especially seen in the F115W imaging, with one ``blob'' in particular extending within the F444W PSF. While the blue component (``blue blob'') peaks in the rest-UV, and thus remains unlikely it would be contributing significantly to the G395M spectrum, we can explicitly test if we are seeing differences driven by varying coverage of the UV-blue components due to the different MSA configurations between programs. We split the extraction aperture in half (spatially) for both C3PO and THRILS observations, and compare the differences in flux of the spectra between each half of the inter- and intra- observations. We create two sub-apertures, approximately half the area of the source slitlet and extract two spectra per source. The extracted spectra are roughly divided along the lines of whether or not they are being dominated by a resolved blue blob. 

If the \oiii\ emission was observed primary from the ``blue blob'', then the total \oiii\ flux in THRILS should be similar to the total \oiii\ flux from C3PO, due to the similar coverage of the blue blob in both slits. Instead we find that the blue blob, which dominates half of the \oiii\ flux in C3PO (see the dotted red spectrum labeled 4 in Figure \ref{fig:subap2}), is roughly equal to the total THRILS \oiii\ flux.  Futhermore, all C3PO sub-apertures are greater than THRILS sub-apertures regardless of ``blue blob'' coverage. While there may be some contribution from the ``blue blob'', an additional flux component is contributing to the total narrow component in the C3PO spectrum, unexplained by aperture differences.





%
\facilities{JWST}

\software{astropy \citep{astropy:2013,astropy:2018},
\jwst\ Pipeline \citep{bushouse22},
          SAOImage DS9 \citep{SmithsonianAstrophysicalObservatory.2000},          }

\bibliography{sample701.bib}
\bibliographystyle{aasjournalv7}

\allauthors

\end{document}